\documentstyle[multicol,epsf,eqsecnum,aps]{revtex}

\begin{document}
\draft
\preprint{WU-HEP-99-5}
\title{Quantum stochastic resonance in driven spin-boson system
with stochastic limit approximation}
\author{%
Kentaro Imafuku,$^{1,}$\thanks{Email: imafuku@mn.waseda.ac.jp}%
Kazuya Yuasa,$^{1,}$\thanks{JSPS Research Fellow. Email:
yuasa@hep.phys.waseda.ac.jp}and
Ichiro Ohba$^{1,2,3,}$\thanks{Email: ohba@mn.waseda.ac.jp}}
\address{%
$^1$Department of Physics, Waseda University, Tokyo 169-8555,
Japan\\
$^2$Advanced Research Center for Science and Engineering, Waseda
University, Tokyo 169-8555, Japan\\
$^3$Kagami Memorial Laboratory for Materials Science and
Technology, Waseda University, Tokyo 169-0051, Japan
}
\date{
September 30, 1999
}
\maketitle
\begin{abstract}
After a brief review of stochastic limit approximation with
spin-boson system from physical points of view, amplification
phenomenon---stochastic resonance phenomenon---in driven
spin-boson system is observed which is helped by the quantum
white noise introduced through the stochastic limit
approximation.
Signal-to-noise ratio resonates at certain temperature if another
noise parameter $\eta$ is chosen properly.
Not only the stochastic resonance in usual sense, but also the
possibilities of the new and interesting
phenomena---``anti-resonance'' and ``double resonance''---are
shown with some choices of $\eta$.
The shift in frequency of the system due to the interaction with
the environment---Lamb shift---has an important role in these
phenomena.
\end{abstract}

\begin{multicols}{2}
\narrowtext
\section{Introduction}
\label{sec:Introduction}
Stochastic resonance (SR) phenomena were first discovered in
connection with periodically recurrent glacial age. Since then
this phenomenon has been found to occur in various fields and has
been attracting wide attention. In short SR is phenomenon
whereby, in contrast to common sense, added noise seems to help
to amplify a signal. Let us briefly review SR phenomenon using
the bistable potential model, driven by a periodic perturbation.
A classical particle in a potential $V(x)$, which has two local
minima, is perturbed by a periodic external force with an
amplitude $\xi$ and a frequency $\Omega$ under the influence of
noise (Fig.~1).
If the amplitude $\xi$ is small, the particle in one of the
stable states cannot go over the potential barrier to the other
stable state [Fig.~1(a)], in other words, the system does not
respond to the input perturbation.
The addition of noise changes the situation; now the
\begin{minipage}{0.48\textwidth}
\bigskip
\begin{figure}
\begin{center}
\epsfxsize=0.94\textwidth
\centerline{\epsfbox{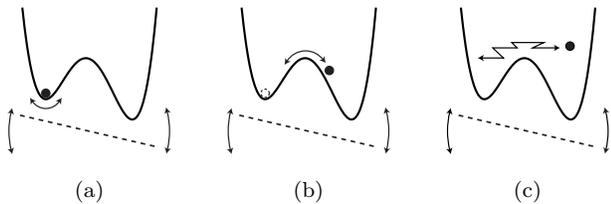}}
\end{center}
\caption{Response of a particle in a bistable potential to an
external periodic perturbation with (a) too small, (b)
appropriate, and (c) too strong noise.}
\label{fig:SR}
\end{figure}
\end{minipage}
particle is kicked by the random force and can go over the
barrier. However if the noise is too strong the response to the
input signal may be smeared. This is because in this case the
particle moves randomly irrespective of the periodicity of the
perturbation [Fig.~1(c)].
However at a certain added noise strength the particle can be
made to travel back and forth between the two stable state, {\em
synchronizing with periodic perturbation of frequency $\Omega$\/}
[Fig.~1(b)].
That is the system responds to the input.

More precisely we can characterize SR as follow:

(1) The power spectrum of the response of a system to a periodic
input has a main sharp peak at the input frequency $\Omega$
\textit{if the noise strength (or temperature) is chosen
properly\/}.

(2) The signal-to-noise ration (SNR) of the response resonates
\textit{at a certain noise strength (or at a certain
temperature)\/}.

Besides the periodicity in the emergence of glacial ages
\cite{ref:Benzi}, SR phenomenon are widely found in nature.
For example SR is found in the nerve of the flagellum of a
crayfish's tail \cite{ref:Crayfish} (however in this case, unlike
the bistable system described above, a threshold reaction is
triggered by noise).
Therefore SR may be a universal concept.

In this paper we discuss SR in bistable model {\em at the quantum
level}, that is quantum stochastic resonance (QSR).
Of course this effect has already been widely studied
\cite{ref:SRReview,ref:QSRReview}, however our particular
interest is ``quantum noise,'' or ``quantum dynamics with
dissipation.''
That is, we are interested in how noise is introduced into the
quantum dynamics to produce QSR.
This is not only important question for QSR, but it is also
relevant for the understanding of several other fundamental
aspects of quantum mechanics, that is the problems of relaxation,
decoherence, measurement and so on.
However quantum mechanics is usually written in terms of causal
deterministic theory governed by unitary time evolution.
It is therefore hard, in principle, to introduce the notion of
``noise'' or ``dissipation'' (with finite degrees of freedom).

In these circumstances there are several different ways to
proceed.
One of the most popular approaches is to introduce an
``environment,'' ``reservoir" or ``heat bath,'' whose detailed
specification one does not know but which has infinite degrees of
freedom.
The whole system (\textit{i.e.}
$\mbox{``system''}+\mbox{``environment''}$) is then treated in
the quantum mechanical way
\cite{ref:FeynmanVernon,ref:CaldeiraLeggett,ref:SpinBoson,ref:Text}.

Through this interaction, the system exchanges energy with the
environment---``dissipation''---, and then it is  reasonable to
assume that some kind of ``noise'' or ``fluctuation'' would
appear due to some ``fluctuation-dissipation relation.''

Along these lines Accardi \textit{et
al.\/}~\cite{ref:Accardi,ref:S-BAccardi,ref:AccardiText} have
introduced the stochastic limit approximation (SLA) as a way to
realize ``quantum white noise.''
The SLA is one way to deal with the van Hove limit, which is the
weak coupling limit given by, $\lambda\to0$ and time
coarse-graining limit given by, $t\mapsto\tau=\lambda^2t$.
This limit ensures that a system in a heat bath approaches
canonical state \cite{ref:Davies}.

As is explicitly shown in Sec.~\ref{sec:SLA} for the spin-boson
system, the spin system in the heat bath composed of bosons
actually approaches the canonical state under the SLA\@.
Furthermore, one can discuss important properties in quantum
dissipative dynamics within this framework, such as the
dependence on temperature of the shift in frequency of the system
due to the interaction with the heat bath.

We here focus our attention on the quantum white noise introduced
through the SLA, and study QSR as part of investigations of the
properties of this noise.
After the introduction of the model to be studied---the driven
spin-boson system---in Sec.~\ref{sec:SBmodel}, the SLA is briefly
reviewed in Sec.~\ref{sec:SLA} with the spin-boson system.
Using this method, we discuss, in Sec.~\ref{sec:QSR}, QSR in the
driven spin-boson system and the role of the quantum fluctuation
and dissipation introduced through the SLA\@.
Section \ref{sec:Summary} is devoted to concluding remarks with
comments on the experimental feasibility of the phenomenon
studied here.
In the Appendix, we add comments on the SLA from a physical point
of view.

\section{Driven Spin-Boson System}
\label{sec:SBmodel}
Here let us introduce the model---the driven spin-boson system
\cite{ref:SRReview,ref:QSRReview}---as a special case of the
bistable model in Sec.~\ref{sec:Introduction}\@.

\noindent
\begin{minipage}{0.48\textwidth}
\begin{figure}
\begin{center}
\epsfxsize=0.94\textwidth
\centerline{\epsfbox{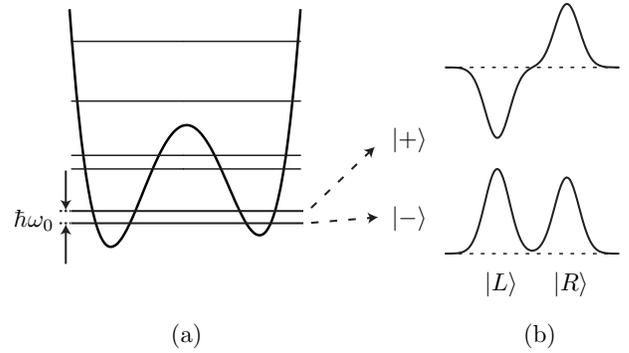}}
\end{center}
\caption{Bistable system at the quantum scale.}
\label{fig:ColdBistable}
\end{figure}
\smallskip
\end{minipage}

\subsection{Spin system}
Consider the situation where the system illustrated in Fig.~2 is
in a deep cold heat bath and its dynamics are ruled mainly by the
lowest tunnel-splitted pair of levels $|\pm\rangle$, where
thermal hopping to upper levels can be neglected.
If the tunneling amplitude between the two wells is sufficiently
small, we are able to consider two ``localized'' states,
$|L\rangle$ and $|R\rangle$, which are approximately the ground
states of left and right wells, respectively.
Taking the set of these states as the Hilbert space basis, this
system can be described by the Hamiltonian,
\begin{equation}
\label{eqn:SystemHamiltonian}
H_S
=\frac{\epsilon}{2}\Bigl(
|R\rangle\langle R|-|L\rangle\langle L|
\Bigr)
+\frac{\Delta}{2}\Bigl(
|R\rangle\langle L|+|L\rangle\langle R|
\Bigr),
\end{equation}
which is essentially the spin-1/2 Hamiltonian with the parameter
$\Delta$ characterizing the tunneling amplitude between the two
wells, and $\epsilon$ characterizing the difference in energy
between the  $|L\rangle$ and $|R\rangle$ states.
Hereafter, we call it the spin system.
By introducing a new basis rotated by the angle
$\theta=\cos^{-1}(\epsilon/\omega_0)=\sin^{-1}(\Delta/\omega_0)$,
\begin{mathletters}
\begin{equation}
|+\rangle
=\cos\frac{\theta}{2}|R\rangle
+\sin\frac{\theta}{2}|L\rangle,
\end{equation}
\begin{equation}
|-\rangle
=-\sin\frac{\theta}{2}|R\rangle
+\cos\frac{\theta}{2}|L\rangle,
\end{equation}
\end{mathletters}
the Hamiltonian $H_S$ is rewritten as a diagonal form
\begin{equation}
\label{eqn:Diagonalized}
H_S
=\frac{\omega_0}{2}\Bigl(
|+\rangle\langle+|-|-\rangle\langle-|
\Bigr).
\end{equation}
The energy gap $\omega_0$ between the two lowest states,
$|+\rangle$ and $|-\rangle$, is given by
\begin{equation}
\omega_0=\sqrt{\epsilon^2+\Delta^2}.
\end{equation}

This two-level system is driven by a periodic forcing with
frequency $\Omega$ and amplitude $\xi$.
This applied force can be described by the perturbative
Hamiltonian
\begin{equation}
\label{eqn:Gaiba}
W=\xi X\sin\Omega t
\end{equation}
with the ``position'' operator $X$ defined by
\begin{equation}
\label{eqn:OrderParameter}
X=|R\rangle\langle R|-|L\rangle\langle L|.
\end{equation}
Of course, there are many other possibilities for the system
driving instead of (\ref{eqn:Gaiba}), e.g.,
$W'=\xi(|R\rangle\langle L|+|L\rangle\langle R|)\sin\Omega t$,
but we choose the perturbation (\ref{eqn:Gaiba}) since it
corresponds to classical SR in the bistable model.

Note that $X$ is an order parameter in discussing QSR phenomenon
in Sec.~\ref{sec:QSR}, which measures the transitions between the
states $|L\rangle$ and $|R\rangle$ under the influence of the
external perturbation.

\subsection{Boson system and its interaction with the spin system}
As mentioned in Sec.~\ref{sec:Introduction}, one must introduce
an ``environment'' for the spin system to dissipate and  be
disturbed.
The environment is chosen as a set of bosons in this paper, whose
Hamiltonian is given by
\begin{equation}
H_B=\int dk\,\omega_ka_k^\dagger a_k.
\end{equation}
Here, $a_k$ and $a_k^\dagger$ are respectively annihilation and
creation operators for a boson of mode $k$ with energy
$\omega_k>0$, and satisfy the commutation relations
\begin{equation}
[a_k,a_{k'}^\dagger]=\delta(k-k'),\quad\mbox{(others)}=0.
\end{equation}

The spin system interacts with the bosons via the interaction
Hamiltonian
\begin{equation}
\label{eqn:Interaction}
\lambda V=\lambda X\int dk\,(g_ka_k^\dagger+g_k^*a_k),
\end{equation}
where $\lambda$ characterizes the strength of the interaction,
and the structure function $g_k$ is a coupling of the bosons of
mode $k$ with the spin system subject to the condition $\int
dk\,|g_k|^2<\infty$.
Note that the spin system and the boson system are coupled with
the bilinear product of the spin operator $X$ and the boson
operators.
Although some specified choices of the coupling may result in
certain outputs, the details of the microscopic Hamiltonian are
not so significant for the derivation of damping dynamics.
A comment on this point can be found in
Sec.~\ref{sec:CouplingChoice}\@.

The system to be analyzed in this paper is thus given by the
total Hamiltonian
\begin{equation}
\label{eqn:TotalHamiltonian}
H=H_0+W+\lambda V,\quad H_0=H_S+H_B.
\end{equation}

\section{Stochastic Limit Approximation}
\label{sec:SLA}
In this section we briefly review the stochastic limit
approximation (SLA) formulated by Accardi \textit{et
al.\/}~\cite{ref:Accardi,ref:S-BAccardi,ref:AccardiText}.
For simplicity, let us consider the case where there is no
external perturbation, i.e., $\xi=0$
\cite{ref:S-BAccardi,ref:AccardiText,ref:SpinBoson}.
The Hamiltonian of the system concerned in this section is thus
\begin{equation}
\label{eqn:S-BHamiltonian}
H_{\mathrm SB}=H_0+\lambda V.
\end{equation}
We entrust the mathematical details to Ref.~\cite{ref:Accardi} or
\cite{ref:AccardiText}, but note that several physically
important points are emphasized and added to the work in
Ref.~\cite{ref:S-BAccardi} and \cite{ref:AccardiText}.
Furthermore in the appendix we add some comments on the SLA taken
from slightly different point of view to that taken by Accardi
\textit{et al.\/}

\subsection{Application to spin-boson system}
In the interaction picture, the time-evolution operator
$U_I^{(\lambda)}\!(t)$ which is governed by the Hamiltonian
(\ref{eqn:S-BHamiltonian}) satisfies the Tomonaga--Schwinger
equation
\begin{mathletters}
\begin{equation}
\frac{d}{dt}U_I^{(\lambda)}\!(t)=-i\lambda
V_I(t)U_I^{(\lambda)}\!(t),\quad
U_I^{(\lambda)}\!(0)=1,
\end{equation}
\begin{equation}
V_I(t)=e^{iH_0t}Ve^{-iH_0t},
\end{equation}
\end{mathletters}
or specifically
\begin{equation}
\label{eqn:InteractionSchroedinger}
\frac{d}{dt}U_I^{(\lambda)}\!(t)
=-i\lambda\sum_\alpha\Bigl(
D_\alpha A_\alpha^\dagger(t)+D_\alpha^\dagger A_\alpha(t)
\Bigr)\,U_I^{(\lambda)}\!(t),
\end{equation}
where $\alpha$ takes $\alpha=\{+,-,0\}$,
\begin{equation}
D_\pm=|\pm\rangle\langle\mp|,\quad
D_0=|+\rangle\langle+|-|-\rangle\langle-|
\end{equation}
are the spin system operators, and
\begin{mathletters}
\begin{equation}
A_\pm(t)
=-\frac{\Delta}{\omega_0}
\int dk\,g^*_ka_ke^{-i(\omega_k\pm\omega_0)t},
\end{equation}
\begin{equation}
A_0(t)=\frac{\epsilon}{\omega_0}
\int dk\,g^*_ka_ke^{-i\omega_kt}
\end{equation}
\end{mathletters}
are the boson system operators.
The SLA is prescribed in the Tomonaga--Schwinger equation
(\ref{eqn:InteractionSchroedinger}) by rescaling time as
$t\mapsto\tau=\lambda^2t$,
\begin{eqnarray}
\label{eqn:RescaledSchroedingerEq}
\frac{d}{d\tau}U_I^{(\lambda)}\!(\tau/\lambda^2)
=&&-i\frac{1}{\lambda}\sum_\alpha\Bigl(
D_\alpha A_\alpha^\dagger(\tau/\lambda^2)
+D_\alpha^\dagger A_\alpha(\tau/\lambda^2)
\Bigr)\,\nonumber\\
&&\qquad{}\times U_I^{(\lambda)}\!(\tau/\lambda^2),
\end{eqnarray}
and then the weak coupling limit $\lambda\rightarrow0$ is taken
(i.e., the van Hove limit \cite{ref:vanHove,ref:Davies}).
As proved in Ref.~\cite{ref:Accardi} or \cite{ref:AccardiText},
there exist the limits
\begin{equation}
\label{eqn:Limits}
\frac{1}{\lambda}A_\alpha(\tau/\lambda^2)
\rightarrow b_\alpha(\tau),\quad
\frac{1}{\lambda}A_\alpha^\dagger(\tau/\lambda^2)
\rightarrow b_\alpha^\dagger(\tau),
\end{equation}
\begin{equation}
U_I^{(\lambda)}\!(\tau/\lambda^2)\rightarrow U_I(\tau),
\end{equation}
and formally
\begin{equation}
\label{eqn:QuantumLangevin}
\frac{d}{d\tau}U_I(\tau)
=-i\sum_\alpha\Bigl(
D_\alpha b^\dagger_\alpha(\tau)+
D_\alpha^\dagger b_\alpha(\tau)
\Bigr)\,U_I(\tau).
\end{equation}
In this limit, the boson operators $b_\alpha(\tau)$ and
$b_\alpha^\dagger(\tau)$ satisfy the commutation relations
\cite{ref:S-BAccardi,ref:AccardiText}
\begin{equation}
\label{eqn:Noise}
[b_-(\tau),b_-^\dagger(\tau')]=2\gamma\delta(\tau-\tau'),\quad
(\mbox{others})=0
\end{equation}
with
\begin{equation}
\label{eqn:Gamma}
\gamma=\left(\frac{\Delta}{\omega_0}\right)^2J(\omega_0),
\end{equation}
\begin{equation}
\label{eqn:SpectralFunction}
J(\omega)=\pi\int dk\,|g_k|^2\delta(\omega_k-\omega).
\end{equation}
For comments on these limits from a physical point of view, see
the Appendix.
The commutation relations (\ref{eqn:Noise}) allow one reasonably
to call $b_\alpha(\tau)$ and $b_\alpha^\dagger(\tau)$ ``quantum
white noise,'' and the Tomonaga--Schwinger equation
(\ref{eqn:QuantumLangevin})  the ``quantum Langevin equation.''
The correlation time is vanishingly small.
However at the same time, we should be careful to note that
equation (\ref{eqn:QuantumLangevin}) is ill-defined.
Fortunately, however, noticing the commutators
\cite{ref:S-BAccardi,ref:AccardiText}
\begin{mathletters}
\label{eqn:[b,U]}
\begin{equation}
[b_\pm(\tau),U_I(\tau)]
=-i\left(\frac{\Delta}{\omega_0}\right)^2
\gamma(\mp\omega_0)D_\pm U_I(\tau),
\end{equation}
\begin{equation}
[b_0(\tau),U_I(\tau)]
=-i\left(\frac{\epsilon}{\omega_0}\right)^2
\gamma(0)D_0U_I(\tau)
\end{equation}
\end{mathletters}
with
\begin{equation}
\label{eqn:RealPart}
\gamma(\omega)=J(\omega)-iI(\omega),\quad
I(\omega)
=\frac{1}{\pi}{\cal P}\!
\int d\omega'\,\frac{J(\omega')}{\omega'-\omega},
\end{equation}
one can evaluate the evolutions of some physically important
quantities.
For example, for the special initial state density operator
\begin{equation}
\rho=\rho_S\otimes\rho_B,\quad
\rho_B=|0\rangle\langle0|
\end{equation}
(i.e., the spin system and the boson system are uncorrelated and
the boson system is in the ground state at $\tau=0$), the
equations for the spin system operators defined by
\begin{equation}
D_\alpha(\tau)
=\mathop{\mathrm tr}\nolimits_B\!\left(
\rho_B\,e^{iH_{\mathrm SB}\tau/\lambda^2}D_\alpha
e^{-iH_{\mathrm SB}\tau/\lambda^2}
\right)
\end{equation}
can be obtained as
\begin{mathletters}
\label{eqn:LangevinVacuum}
\begin{equation}
\frac{d}{d\tau}D_\pm(\tau)=-(\gamma\mp i\omega_R)D_\pm(\tau),
\end{equation}
\begin{equation}
\frac{d}{d\tau}D_0(\tau)=-2\gamma D_0(\tau)-2\gamma,
\end{equation}
\end{mathletters}
which give the exponentially decaying dynamics
\begin{mathletters}
\label{eqn:SolutionVacuum}
\begin{equation}
D_\pm(\tau)=D_\pm e^{-(\gamma\mp i\omega_R)\tau},
\end{equation}
\begin{equation}
D_0(\tau)=(D_0+1)e^{-2\gamma\tau}-1.
\end{equation}
\end{mathletters}
Here $\omega_R$ is the renormalized frequency
\begin{equation}
\label{eqn:RenormalizedFrequency}
\omega_R
=\omega_0/\lambda^2-\sigma
=\tilde{\omega}_0-\sigma,
\end{equation}
where the frequency shift $\sigma$ emerges due to the interaction
\begin{equation}
\label{eqn:LambShift}
\sigma=\left(\frac{\epsilon}{\omega_0}\right)^2
\Bigl(I(\omega_0)-I(-\omega_0)\Bigr).
\end{equation}
Note that $\mathop{\mathrm tr}_B$ denotes the trace over the
boson-degrees of freedom. This is the procedure for ``partial
trace.''
It reduces the effects of the interaction between the spin system
and the boson environment to the spectral function $J(\omega)$
defined in Eq.~(\ref{eqn:SpectralFunction}).
The damping coefficient $\gamma$ in Eq.~(\ref{eqn:Gamma}) and the
frequency shift $\sigma$ in Eq.~(\ref{eqn:LambShift}) with
Eq.~(\ref{eqn:RealPart}) are both given in terms of $J(\omega)$.

It is also possible to evaluate $\gamma$ and $\sigma$ for the
boson environment at finite temperature $T$
\begin{equation}
\rho_B
=e^{-\beta H_B}/\mathop{\mathrm tr}\nolimits_Be^{-\beta H_B}
\end{equation}
by using the TFD technique \cite{ref:TFD}, for example.
Here $\beta=1/k_BT$ with $k_B$ being the Boltzmann constant.
In this case, one obtains
\begin{mathletters}
\label{eqn:SolutionBeta}
\begin{equation}
D_\pm(\tau)
=D_\pm e^{-(\gamma^\beta\mp i\omega_R^\beta)\tau},
\end{equation}
\begin{equation}
D_0(\tau)=
\left(
D_0+\frac{\gamma}{\gamma^\beta}
\right)e^{-2\gamma^\beta\tau}
-\frac{\gamma}{\gamma^\beta}
\end{equation}
\end{mathletters}
with the temperature affected parameters
\begin{mathletters}
\label{eqn:ParsBeta}
\begin{equation}
\label{eqn:DissipationConstantsBeta}
\gamma^\beta
=\left(\frac{\Delta}{\omega_0}\right)^2J^\beta\!(\omega_0),
\end{equation}
\begin{equation}
\omega_R^\beta
=\omega_0/\lambda^2-\sigma^\beta
=\tilde{\omega}_0-\sigma^\beta,
\end{equation}
\begin{equation}
\label{eqn:LambShiftBeta}
\sigma^{\beta}
=\left(\frac{\epsilon}{\omega_0}\right)^2
\Bigl(I^\beta\!(\omega_0)-I^\beta\!(-\omega_0)\Bigr),
\end{equation}
\end{mathletters}
and the functions
\begin{mathletters}
\label{eqn:FuncsBeta}
\begin{equation}
\label{eqn:JBeta}
J^\beta\!(\omega)=J(\omega)\coth\frac{1}{2}\beta\omega,
\end{equation}
\begin{equation}
\label{eqn:IBeta}
I^\beta\!(\omega)
=\frac{1}{\pi}{\cal P}\!
\int d\omega'\,\frac{J^\beta\!(\omega')}{\omega'-\omega}.
\end{equation}
\end{mathletters}
The damping coefficient $\gamma^\beta$ and the frequency shift
$\sigma^\beta$ are obtained from $\gamma$ and $\sigma$,
respectively, by replacing the spectral function $J(\omega)$ with
the temperature modified one $J^\beta\!(\omega)$.

Notice that the long-time limits of the operators
$D_0(\tau)\rightarrow-\tanh(\beta\omega_0/2)$ and
$D_\pm(\tau)\rightarrow0$ are both c-numbers (or unit operators
of the spin system multiplied by c-numbers).
This means that the spin system approaches some unique state
irrespective of the initial state $\rho_S$.
In fact, the averages of any spin system operators, which are
composed of $D_\alpha(\tau)$, approach unique values.
One can further confirm that the long-time limit of the state of
the spin system is nothing but the thermal state at the
temperature $T$.
Taking averages of $D_\alpha(\tau)$ with some arbitrary initial
state $\rho_S$, one obtains the matrix elements of the system
density operator defined by
\begin{equation}
\rho_S(\tau)
=\mathop{\mathrm tr}\nolimits_B\rho(\tau),
\end{equation}
\begin{equation}
\rho(\tau)
=e^{-iH_{\mathrm SB}\tau/\lambda^2}\rho\,
e^{iH_{\mathrm SB}\tau/\lambda^2}.
\end{equation}
Their dynamics are immediately obtained from the equations
(\ref{eqn:SolutionBeta}), and their long-time limits are given by
\begin{mathletters}
\label{eqn:THState}
\begin{equation}
\label{eqn:THStateNonDiagonal}
\langle-|\rho_S(\tau)|+\rangle
=\langle D_+(\tau)\rangle
\rightarrow0,
\end{equation}
\begin{eqnarray}
\langle\pm|\rho_S(\tau)|\pm\rangle
&&=\frac{1}{2}\Bigl(1\pm\langle D_0(\tau)\rangle\Bigr)\nonumber\\
&&\rightarrow\frac{e^{\mp\beta\omega_0/2}}%
{e^{\beta\omega_0/2}+e^{-\beta\omega_0/2}},
\label{eqn:THStateDiagonal}
\end{eqnarray}
\end{mathletters}
which are equivalent to
\begin{equation}
\rho_S(\tau)
\rightarrow
e^{-\beta H_S}/\mathop{\mathrm tr}\nolimits_Se^{-\beta H_S},
\end{equation}
i.e., the system approaches the thermal equilibrium state at
temperature $T$ through decoherence
(\ref{eqn:THStateNonDiagonal}).
Here $\mathop{\mathrm tr}\nolimits_S$ denotes the trace over the
spin-degrees of freedom.

\subsection{Comments from physical points of view}
\label{sec:Comments}
\subsubsection{Orders of parameters}
\label{sec:Orders}
It is important to clarify the order of magnitude of the
parameters.
In this formalism, one considers that the new time $\tau$ is
physical and that, if they are measured in this macroscopic time,
the parameters of the spin system should have some meaningful
values, e.g., $\omega_R^\beta$ or $\tilde{\omega}_0$, instead of
$\omega_0$, should be finite.
On the other hand, the time scales of the boson system should be
measured in the microscopic time $t$.
This can be seen in the emergence of the delta function in
Eq.~(\ref{eqn:Noise}).
This is due to the coarse-graining in time,
$t\mapsto\tau=\lambda^2t\,(\lambda\rightarrow0)$.
It reflects the fact that characteristic time scales of the boson
system, like the correlation time for example, are vanishingly
small when measured in the macroscopic time.  That is they are
negligible when compared to the characteristic  times of the spin
system, such as $1/\tilde{\omega}_0$ for example.
This is the situation which occurs in the stochastic limit.

As for the temperature, the physically interesting situation is
where the temperature $T$ is such that $\beta\omega_0$ has some
finite value, which is contained in $\gamma^\beta$ and
$\sigma^\beta$ through $J^\beta\!(\omega_0)$ [see
Eqs.~(\ref{eqn:ParsBeta}) and (\ref{eqn:FuncsBeta})] and in the
thermal equilibrium distribution (\ref{eqn:THStateDiagonal}).
That is, the rescaled temperature $\tilde{T}=T/\lambda^2$ should
be finite.

\subsubsection{Choice of the coupling}
\label{sec:CouplingChoice}
In this paper, the spin-boson system is coupled through the
specific choice of the coupling, in particular, the choice $X$.
Of course, there are many other possibilities, such as
$V=(|R\rangle\langle L|+|L\rangle\langle R|)\int
dk\,(g_ka_k^\dagger+g_k^*a_k)$, but they all result in the same
damping dynamics (\ref{eqn:SolutionBeta}) except for the
prefactors $(\Delta/\omega_0)^2$ and $(\epsilon/\omega_0)^2$ in
the definitions of $\gamma^\beta$ and $\sigma^\beta$ in
Eqs.~(\ref{eqn:DissipationConstantsBeta}) and
(\ref{eqn:LambShiftBeta}).
[One has to notice, however, that the special choice of the
interaction $V=(|+\rangle\langle+|-|-\rangle\langle-|)\int
dk\,(g_ka_k^\dagger+g_k^*a_k)$ gives the special situation
$\gamma^\beta=0$, which is also included in
Eqs.~(\ref{eqn:SolutionBeta}).
The possibility of $\sigma^\beta=0$ also exists for some special
choices of $V$.]
From a semi-phenomenological point of view,
Eqs.(\ref{eqn:SolutionBeta}) are sufficient for the description
of experiments.
Knowledge of the parameters $\tilde{\omega}_0$, $\gamma^\beta$,
and $\sigma^\beta$ from an experiment would enable one to predict
the dynamical development of any quantities.
Details of the microscopic Hamiltonian would not have much effect
on the macroscopic behavior.

\section{Quantum Stochastic Resonance}
\label{sec:QSR}
Now let us discuss QSR in the driven spin-boson system
(\ref{eqn:TotalHamiltonian}) with the SLA\@.
By observing the response of the system to the external
perturbation (\ref{eqn:Gaiba}) through the dynamics of the
``position'' operator
\begin{equation}
X(\tau)=\mathop{\mathrm tr}\nolimits_B\!\left(
\rho_B\,e^{iH\tau/\lambda^2}Xe^{-iH\tau/\lambda^2}
\right),
\end{equation}
which measures the transitions of the system between the left
state $|L\rangle$ and the right state $|R\rangle$, we see an
amplification of the input external perturbation with the
addition of noise.

The Tomonaga--Schwinger equation is now
\begin{mathletters}
\begin{equation}
\label{eqn:PerturbedSchroedingerEq}
\frac{d}{dt}U_I^{(\lambda)}\!(t)
=-i\Bigl(\lambda V_I(t)+W_I(t)\Bigr)\,
U_I^{(\lambda)}\!(t),\quad
U_I^{(\lambda)}\!(0)=1,
\end{equation}
\begin{equation}
W_I(t)=\xi\left[
\frac{\epsilon}{\omega_0}D_0
-\frac{\Delta}{\omega_0}\left(
D_+e^{i\omega_0t}+D_-e^{-i\omega_0t}
\right)
\right]\sin\Omega t,
\end{equation}
\end{mathletters}
and along the same lines as in Sec.~\ref{sec:SLA}, the SLA
($t\mapsto\tau=\lambda^2t$, $\lambda\rightarrow0$) is taken to
give the ``quantum Langevin equation''
\begin{eqnarray}
\frac{d}{d\tau}U_I(\tau)
=&&-i\sum_\alpha\Bigl(
D_\alpha b_\alpha^\dagger(\tau)+D_\alpha^\dagger b_\alpha(\tau)
\Bigr)U_I(\tau)\nonumber\\
&&{}-i\tilde{\xi}\left[
\frac{\epsilon}{\omega_0}D_0
-\frac{\Delta}{\omega_0}\left(
D_+e^{i\tilde{\omega}_0\tau}
+D_-e^{-i\tilde{\omega}_0\tau}
\right)
\right]\nonumber\\
&&\nonumber\\
&&\qquad{}\times\sin(\tilde{\Omega}\tau)U_I(\tau).
\label{eqn:PerturbedLangevin}
\end{eqnarray}
Note that the parameters are rescaled as
$\tilde{\omega}_0=\omega_0/\lambda^2$,
$\tilde{\Omega}=\Omega/\lambda^2$, and
$\tilde{\xi}=\xi/\lambda^2$ according to time rescaling, and are
assumed to take physical values if measured in the macroscopic
time.
The equations of the spin system operators $D_\alpha(\tau)$ are
then given by
\begin{mathletters}
\begin{eqnarray}
&&\frac{d}{d\tau}D_\pm(\tau)
=-(\gamma^\beta\mp i\omega_R^\beta)
D_\pm(\tau)\nonumber\\
&&\qquad{}\pm2i\tilde{\xi}\left(
\frac{\epsilon}{\omega_0}D_\pm(\tau)
+\frac{1}{2}\frac{\Delta}{\omega_0}D_0(\tau)
\right)\sin\tilde{\Omega}\tau,
\end{eqnarray}
\begin{eqnarray}
&&\frac{d}{d\tau}D_0(\tau)
=-\gamma^\beta D_0(\tau)-\gamma\nonumber\\
&&\qquad{}+2i\tilde{\xi}
\frac{\Delta}{\omega_0}\Bigl(
D_+(\tau)-D_-(\tau)
\Bigr)\sin\tilde{\Omega}\tau.
\end{eqnarray}
\end{mathletters}
These are, however, difficult to solve exactly, so we rely upon
the perturbation method and assume that  the external
perturbation is weak.
(One is interested here in SR phenomenon, i.e., amplification of
{\em weak} inputs with the help of noise.)
The solutions up to $O(\tilde{\xi}/\tilde{\omega}_0)$ or
$O(\xi/\omega_0)$ are thus obtained for long times
$\tau\gg1/\gamma^\beta$ as
\begin{mathletters}
\begin{eqnarray}
D_\pm(\tau)
\rightarrow&&\mp\frac{i}{2}\tilde{\xi}\frac{\Delta}{\omega_0}
\tanh\frac{1}{2}\beta\omega_0
\left[
\frac{1}{(\omega_R^\beta-\tilde{\Omega})
\pm i\gamma^\beta}
e^{\pm i\tilde{\Omega}\tau}
\right.\nonumber\\
&&\qquad\left.{}-\frac{1}{(\omega_R^\beta+\tilde{\Omega})
\pm i\gamma^\beta}
e^{\mp i\tilde{\Omega}\tau}
\right],
\end{eqnarray}
\begin{equation}
\label{eqn:StationarySolution}
D_0(\tau)
\rightarrow-\tanh\frac{1}{2}\beta\omega_0,
\end{equation}
\end{mathletters}
and $X(\tau)$, by combining these solutions, as
\begin{eqnarray}
&&X(\tau)
=\frac{\epsilon}{\omega_0}D_0(\tau)
-\frac{\Delta}{\omega_0}\Bigl(D_+(\tau)+D_-(\tau)\Bigr)
\nonumber\\
&&\qquad
{}\rightarrow-\frac{\epsilon}{\omega_0}
\tanh\frac{1}{2}\beta\omega_0
{}+\frac{\xi}{\omega_0}
A^\beta\!(\tilde{\Omega})\sin\!\left(
\tilde{\Omega}\tau-\phi^\beta\!(\tilde{\Omega})
\right).\nonumber\\
\label{eqn:Response}
\end{eqnarray}
Here the amplitude $A^\beta\!(\tilde{\Omega})$ and the phase
delay $\phi^\beta\!(\tilde{\Omega})$ are given, respectively, by
\begin{equation}
\label{eqn:Amplitude}
A^\beta\!(\tilde{\Omega})
=\frac{
2(\Delta/\omega_0)^2\tilde{\omega}_0\omega_R^\beta
\tanh(\beta\omega_0/2)
}{
\sqrt{\Bigl(
(\omega_R^\beta)^2-\tilde{\Omega}^2+(\gamma^\beta)^2
\Bigr)^2+\Bigl(2\gamma^\beta\tilde{\Omega}\Bigr)^2}
}
\end{equation}
and
\begin{equation}
\label{eqn:PhaseDelay}
\tan\phi^\beta\!(\tilde{\Omega})
=\frac{2\gamma^\beta\tilde{\Omega}}{
(\omega_R^\beta)^2-{\tilde{\Omega}}^2+(\gamma^\beta)^2
}.
\end{equation}
Responding to the input perturbation, $X(\tau)$ oscillates around
the thermal equilibrium state with the frequency of the
perturbation $\tilde{\Omega}$.

Let us define the signal-to-noise ratio (SNR)
$R^\beta\!(\tilde{\Omega})$ by
\begin{equation}
\label{eqn:SNR}
R^\beta\!(\tilde{\Omega})
=|A^\beta\!(\tilde{\Omega})|/(\gamma^\beta\!/\tilde{\omega}_0),
\end{equation}
and study its dependence on the temperature.
If this resonates at any certain temperature we conclude SR
exists in this model.
In the following, we show the analyses for two specific choices
of the spectral function $J(\omega)$, as examples.
One choice is
\begin{mathletters}
\label{eqn:ModelSpectralFunction}
\begin{equation}
\label{eqn:Ohmic}
J(\omega)=\left\{
\begin{array}{ll}
\displaystyle
\eta\omega
&\displaystyle
(0<\omega<\Lambda)\\
\displaystyle\\
\displaystyle
\eta\Lambda
&\displaystyle
(\omega>\Lambda)
\end{array}
\right.,
\end{equation}
called here the ``Ohmic case'' [Fig.~3(b)], and the other is

\noindent
\begin{minipage}{0.48\textwidth}
\bigskip
\begin{figure}
\begin{center}
\epsfxsize=0.75\textwidth
\centerline{\epsfbox{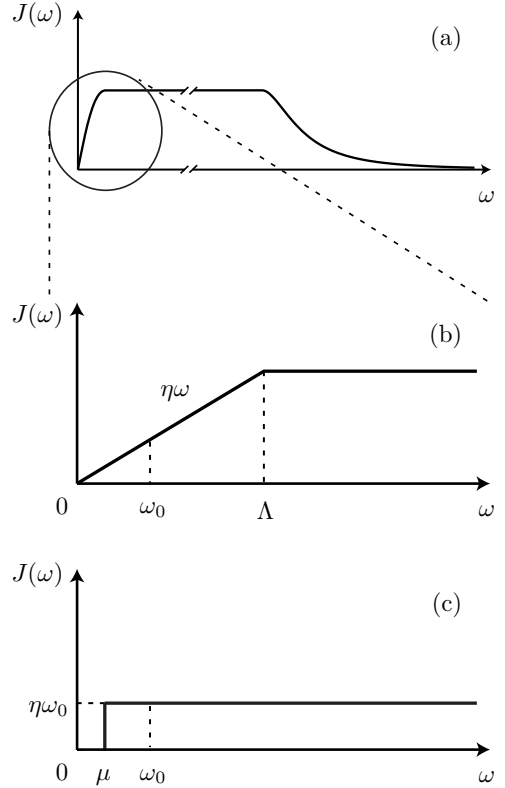}}
\end{center}
\caption{Functional form of (a) physically realistic, (b) ``Ohmic
case,'' and (c) ``constant case'' spectral function $J(\omega)$.}
\label{fig:SpectralFunction}
\end{figure}
\end{minipage}
\begin{equation}
\label{eqn:Constant}
J(\omega)=\left\{
\begin{array}{ll}
\displaystyle
0
&\displaystyle
(0<\omega<\mu)\\
\displaystyle\\
\displaystyle
\eta\omega_0
&\displaystyle
(\omega>\mu)
\end{array}
\right.,
\end{equation}
\end{mathletters}
called the ``constant case''  [Fig.~3(c)].
Although a physically realistic spectral function $J(\omega)$ may
have a cutoff at high frequency $\Lambda_C$ as sketched in
Fig.~3(a), it may be reasonable to consider that $\Lambda_C$ can
be infinitely large compared to the characteristic frequency of
the spin system $\omega_0$ in the stochastic limit situation.
We hence adopt the model spectral functions given by
Eqs.~(\ref{eqn:Ohmic}) and (\ref{eqn:Constant}) and illustrated
in Figs.~3(b) and (c).
Their names come from the functional forms in the regions around
``on-shell'' $\omega=\omega_0$ which are assumed to be the
regions $0<\omega<\Lambda$ and $\mu<\omega$ for each case.
Note that $J(\omega_0)$, and hence $\gamma^\beta$ given by
Eq.~(\ref{eqn:DissipationConstantsBeta}), have the same value for
both cases, and the dimensionless parameter $\eta$ controls its
magnitude, i.e., the noise strength.
The difference between the two cases manifests itself in the
temperature dependence of the frequency shift $\sigma^\beta$
given by Eq.~(\ref{eqn:LambShiftBeta}) (Fig.~4).
Note further that the function $I^\beta\!(\omega)$ defined by the
dispersion relation (\ref{eqn:IBeta}) does not converge with the
model spectral functions (\ref{eqn:ModelSpectralFunction}).
Since the asymptotic behaviors of $J^\beta\!(\omega)$ in these
models are constant as $\omega\to\infty$, one has to apply a
subtracted form to the dispersion relation.
After a subtraction at $\omega=\omega_1$, this becomes
\begin{equation}
\label{eqn:Subtracted}
I^\beta\!(\omega)
=I^\beta\!(\omega_1)
+\frac{\omega-\omega_1}{\pi}{\cal P}\!\int d\omega'\,
\frac{J^\beta\!(\omega')-J^\beta\!(\omega_1)}%
{(\omega'-\omega_1)(\omega'-\omega)}.
\end{equation}
Choosing the subtraction point as $\omega_1=0$, one gets a
convergent integral,
\begin{equation}
\label{eqn:ZeroSubtracted}
I^\beta\!(\omega)
=I^\beta\!(0)
+\frac{\omega}{\pi}{\cal P}\!\int d\omega'\,
\frac{J^\beta\!(\omega')}{\omega'(\omega'-\omega)}.
\end{equation}
From Eqs.~(\ref{eqn:IBeta}) and (\ref{eqn:ZeroSubtracted}), the
frequency shift is given by
\begin{equation}
\sigma^\beta=\left(\frac{\epsilon}{\omega_0}\right)^2
\frac{2\omega_0}{\pi}{\cal P}\!\int d\omega'\,
\frac{J^\beta\!(\omega')}{\omega'^2-\omega_0^2}.
\end{equation}

There are two important parameters concerning the environment or
the noise, i.e., the temperature $T$ and
\begin{minipage}{0.48\textwidth}
\bigskip
\begin{figure}
\begin{center}
\epsfxsize=0.96\textwidth
\centerline{\epsfbox{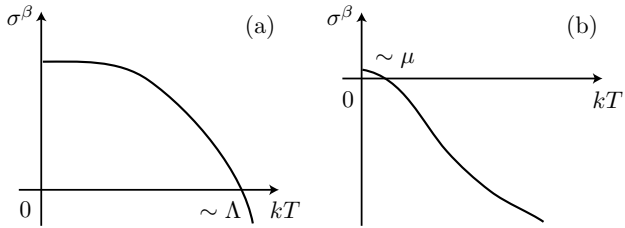}}
\end{center}
\caption{Schematic forms of the frequency shift for (a) ``Ohmic
case'' and (b) ``constant case.''}
\label{fig:LambShift}
\end{figure}
\end{minipage}
\begin{minipage}{0.48\textwidth}
\begin{figure}
\begin{center}
\epsfxsize=\textwidth
\centerline{\epsfbox{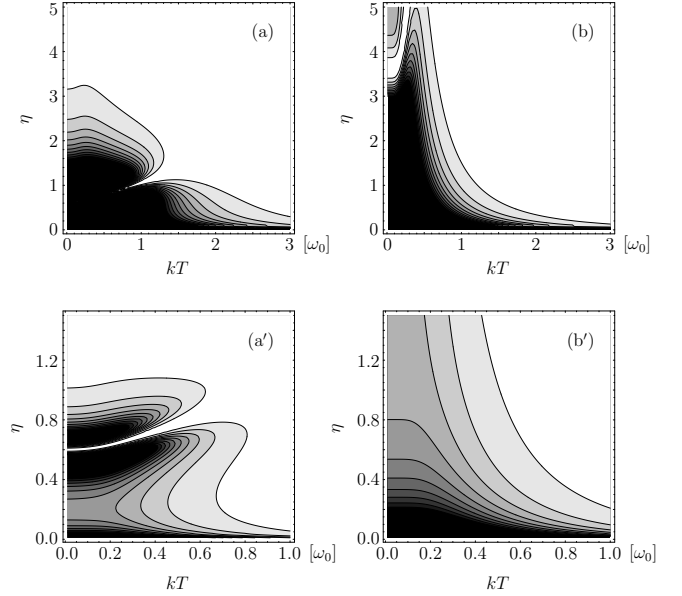}}
\end{center}
\caption{Temperature- and $\eta$-dependence of SNR for (a)
``Ohmic case'' and (b) ``constant case'' with
$\Omega=0.10\omega_0$, $\Lambda=2.0\omega_0$, $\mu=0.50\omega_0$,
and $\Delta/\omega_0=0.35$.
(a$^\prime$) and (b$^\prime$) are enlarged versions of (a) and
(b), respectively.
Darker grays correspond to larger $R^\beta\!(\tilde{\Omega})$.}
\label{fig:Contour}
\end{figure}
\medskip
\end{minipage}
the noise strength $\eta$.
In Fig.~5, the SNRs $R^\beta\!(\tilde{\Omega})$ for both cases
are shown in the $\eta$-$T$ plane.
One may realize at first sight that SNR depends deeply on the
choice of $J^\beta\!(\omega)$, namely, on the temperature
dependence of the frequency shift $\sigma^\beta$.
The temperature dependences of the SNRs are shown in Fig.~6(a)
for the ``Ohmic case'' with $\eta=0.59$, and in Fig.~6(b) for the
``constant case'' with $\eta=3.5$.
maximum values are seen at around the temperature
$kT\sim0.3\omega_0$ for both cases, that is, SR occurs.
Roughly speaking, these maximum points correspond to the minima
of the denominator in the right hand side of
Eq.~(\ref{eqn:Amplitude}).
One has to notice, however, that this does not occur for all
$\eta$: for some $\eta$ it occurs, and for others it does not.
And beyond these two possibilities, one can find some strange
phenomena.
See Fig.~6(a$^\prime$), where $\eta$ is chosen as $\eta=0.65$ for
the ``ohmic case,'' and Fig.~6(b$^\prime$), where $\eta=4.5$ for
the ``constant case.''
There exist temperatures where the system does not respond.
We may call this ``anti-resonance.''
It occurs when the frequency shift $\sigma^\beta$ coincides with
the system frequency $\tilde{\omega}_0$.
See the numerator of the amplitude $A^\beta\!(\tilde{\Omega})$.
And see Fig.~6(a$^{\prime\prime}$) with $\eta=0.70$ for the
``ohmic case'' and Fig.~6(b$^{\prime\prime}$) with $\eta=15$ for
the ``constant case,'' where the SNRs have two peaks, i.e.,
``double resonance.''
The second maximum comes from the overlapping effect of a
negatively decreasing factor $\omega_R^\beta$ beyond its zero
point and a positively decreasing Planck distribution.
And it does not correspond to a genuine SR.
It has no counterpart in classical systems.
The behavior of the frequency shift $\sigma^\beta$ may
\begin{minipage}{0.48\textwidth}
\begin{figure}
\begin{center}
\epsfxsize=\textwidth
\centerline{\epsfbox{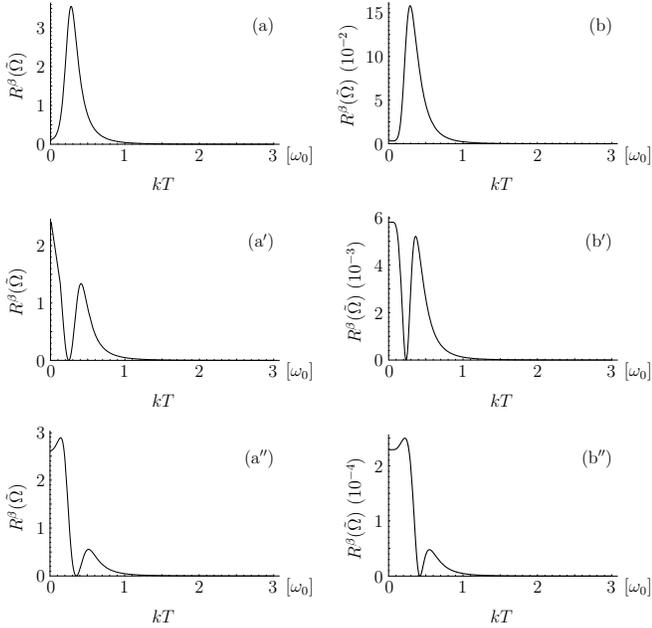}}
\end{center}
\caption{Temperature-dependence of SNR for ``Ohmic case'' with
(a) $\eta=0.59$, (a$^\prime$) $\eta=0.65$, and
(a$^{\prime\prime}$) $\eta=0.70$, and for ``constant case'' with
(b) $\eta=3.5$, (b$^\prime$) $\eta=4.5$, and (b$^{\prime\prime}$)
$\eta=15$.}
\label{fig:Temperature}
\end{figure}
\bigskip
\end{minipage}
be the key to these phenomena.

\section{Summary}
\label{sec:Summary}
QSR in the driven spin-boson system is discussed with  quantum
white noise introduced through the SLA\@.
The SLA is a framework which can be used to describe the van Hove
limit, which ensures the approach of system in a thermal
environment to the thermal equilibrium state.
SNRs versus noise parameters---temperature $T$ and the noise
strength $\eta$---are studied with two model spectral functions
$J(\omega)$.
The occurrence of SR depends on the choice of $\eta$.
For some $\eta$ the system does not resonate, for other values it
does, and a new phenomenon---anti-resonance and double
resonance--- is observed.
The temperature dependence of the frequency shift of the system
$\sigma^\beta$ due to the interaction with the environment---Lamb
shift---may be the key to these phenomena.
In this sense, QSR owes its existence to a quantum effect, which
is different from the classical SR, where random force itself is
important.
To understand this point, this system should be studied in the
crossover area between the quantum and classical regimes.
This work is now in progress.

It should further be emphasized that the analysis here is from
the microscopic view point, not from a semi-phenomenological
viewpoint. In the latter there is no criterion which would tell
us how to incorporate the damping coefficient $\gamma^\beta$ and
the frequency shift $\sigma^\beta$ into the phenomenological
equation properly.
Here the damping dynamics is obtained from the fundamental
microscopic Hamiltonian underneath the theory.

Finally, let us mention experimental situations for the present
analysis.

(1) The physical time is not $t$ but $\tau$. Experimental data
should be compared with the theoretical predictions from the
analysis in this paper in the macroscopic time $\tau$.

(2) It is very difficult in general to prepare precise quantum
mechanical initial conditions experimentally.
Fortunately, however, SNR is obtained from the stationary
behavior of the system at large times $\tau\gg1/\gamma^\beta$,
and is irrespective of the initial condition.

(3) It is possible to control $J(\omega)$ in the cavity QED
experiment.
In fact, the life-time of the unstable state of an atom can be
successfully controlled by changing the modes of the
electromagnetic field, i.e., by changing $J(\omega)$.
This means that it may be possible to observe SNRs with different
choices of $J(\omega)$ in the cavity QED.

There may be technical difficulties to overcome, but it may be
possible to observe experimentally the phenomena predicted here.
This would  also be an experimental verification of SLA itself.

\acknowledgments
The authors acknowledge helpful and fruitful discussions with
Profs.~H.~Nakazato and S.~Pascazio.
They also thank Profs.~L.~Accardi, I.~V.~Volovich, and N.~Obata for discussions on the stochastic limit approximation, Prof.~C.~Uchiyama for discussions at JPS meetings, and Prof.~H.~Hasegawa for discussions after RIMS meeting.

This work is supported partially by Grant-in-Aid for JSPS
Research Fellows and Waseda University Grant for Special Research
Project.

\appendix
\section*{}
\label{app:SLA}
Here we briefly describe SLA from a physical point of view.
Introducing a generalized rescaled time as
\begin{equation}
t\mapsto\tau=\lambda^\nu t,\quad\nu>0,
\end{equation}
one has the Tomonaga--Schwinger equation
\begin{eqnarray}
\label{eqn:GenerallyRescaledEquation}
\frac{d}{d\tau}U_I^{(\lambda)}\!(\tau/\lambda^\nu)
=&&-i\frac{1}{\lambda^{\nu-1}}\sum_\alpha\Bigl(
D_\alpha A_\alpha^\dagger(\tau/\lambda^\nu)\nonumber\\
&&\qquad{}+D_\alpha^\dagger A_\alpha(\tau/\lambda^\nu)
\Bigr)\,U_I^{(\lambda)}\!(\tau/\lambda^\nu).
\end{eqnarray}
We require that the rescaled operators $A_\alpha$,
$A_\alpha^\dag$ should satisfy the commutation relations with
respect to the rescaled time $\tau$.
It is easily shown that the possible choices are only of the form
$A_\alpha(\tau/\lambda^\nu)/\lambda^{\nu/2}$ and that the
non-trivial commutation relation is
\begin{eqnarray}
&&\left[
\frac{1}{\lambda^{\nu/2}}A_-(\tau/\lambda^\nu),
\frac{1}{\lambda^{\nu/2}}A_-^\dag(\tau'/\lambda^\nu)
\right]\nonumber\\
&&\qquad=2\left(\frac{\Delta}{\omega_0}\right)^2\left[
J(\omega_0)
+i\lambda^\nu J'(\omega_0)\frac{\partial}{\partial\tau}
+\cdots
\right]\nonumber\\
&&\qquad\qquad\qquad\qquad\qquad
{}\times\frac{1}{2\pi}\int_{-\omega_0/\lambda^\nu}^\infty\!dx\,
e^{-ix(\tau-\tau')},
\end{eqnarray}
while the others vanish.
If the first and higher derivatives of the spectral function
$J^{(n)}(\omega_0)\,(n=1,2,\ldots)$ do not have singularities,
one can safely neglect all terms other than $J(\omega_0)$ from
the expansion.
This corresponds simply to the choice of diagonal singularity,
i.e., only the boson mode $\omega_k=\omega_0$ contributes to the
damping coefficient in the scaling limit $\lambda\to0$.

From the above considerations, it is convenient to rewrite
Eq.~(\ref{eqn:GenerallyRescaledEquation}) as
\begin{eqnarray}
\label{eqn:RewrittenGenerallyRescaledEquation}
&&\frac{d}{d\tau}U_I^{(\lambda)}\!(\tau/\lambda^\nu)\nonumber\\
&&\qquad=-i\lambda^{1-\nu/2}\sum_\alpha\biggl(
D_\alpha\frac{1}{\lambda^{\nu/2}}
A_\alpha^\dagger(\tau/\lambda^\nu)\nonumber\\
&&\qquad\qquad\qquad
{}+D_\alpha^\dagger\frac{1}{\lambda^{\nu/2}}
A_\alpha(\tau/\lambda^\nu)
\biggr)\,U_I^{(\lambda)}\!(\tau/\lambda^\nu).
\end{eqnarray}
Thus, one can see that, in the limit $\lambda\to0$,
(1) the right hand side of
Eq.~(\ref{eqn:RewrittenGenerallyRescaledEquation}) vanishes in
the case of ``under'' SLA ($\nu<2$), while
(2) it diverges in the case of ``over'' SLA ($\nu>2$), and
(3) it has a formal limit in the case of ``critical'' SLA
($\nu=2$).
Therefore the only meaningful result occurs in the $\nu=2$ case.


\end{multicols}


\begin{references}
\bibitem{ref:SRReview}
For reviews, see L. Gammaitoni, P. H\"anggi, P. Jung, and
F. Marchesoni, Rev.\ Mod.\ Phys.\ {\bf70}, 223 (1998).
\bibitem{ref:Benzi}
R. Benzi, G. Parisi, A. Sutera, and A. Vulpiani, Tellus {\bf34},
10 (1982).
\bibitem{ref:Crayfish}
J. K. Douglass, L. Wilkens, E. Pantazelou, and F. Moss, Nature
{\bf365}, 337 (1993).
\bibitem{ref:QSRReview}
For reviews, see M. Grifoni and P. H\"anggi, Phys.\ Rep.\ {\bf
304}, 229 (1998).
They discussed the system described by
Eq.~(\ref{eqn:TotalHamiltonian}), based on the approach quoted in
the paper by Leggett \textit{et al.\/}~[A.\ J.\ Leggett
\textit{et al.\/}, Rev.\ Mod.\ Phys.\ {\bf59}, 1 (1987)].
They also obtained the non-linear response of an order parameter
to an applied field, and their works have given us many
stimulations.
But it is hard to get result analytically except for some special
parameter.
\bibitem{ref:FeynmanVernon}
R. P. Feynman and F. L. Vernon, Ann.\ Phys.\ \textbf{24}, 118
(1963);
R. P. Feynman and A. R. Hibbs, \textit{Quantum Mechanics and Path
Integrals\/} (McGraw-Hill, New York, 1965).
\bibitem{ref:CaldeiraLeggett}
A. O. Caldeira and A. J. Leggett, Ann.\ Phys.\ \textbf{149}, 374
(1983); \textbf{153}, 445(E) (1984);
Physica \textbf{121A}, 587 (1983); \textbf{130A}, 374(E) (1985).
\bibitem{ref:SpinBoson}
A. J. Leggett, S. Chakravarty, A. T. Dorsey, M. P. A. Fisher,
A. Garg, and W. Zwerger, Rev.\ Mod.\ Phys.\ \textbf{59}, 1
(1987).
\bibitem{ref:Text}
U. Weiss, \textit{Quantum Dissipative Systems\/}, Vol.\ 2 of
\textit{Series in Modern Condensed Matter Physics\/} (World
Scientific, Singapore, 1993).
\bibitem{ref:Accardi}
L. Accardi, A. Frigerio, and Y. G. Lu, Commun.\ Math.\ Phys.\
{\bf131}, 537 (1990);
L. Accardi, J. Gough, and Y. G. Lu, Rep.\ Math.\ Phys.\ {\bf36},
155 (1995).
\bibitem{ref:S-BAccardi}
L. Accardi, S. V. Kozyrev, and I. V. Volovich, Phys.\ Rev.\ A
{\bf56}, 2557 (1997).
\bibitem{ref:AccardiText}
L. Accardi, Y. G. Lu, and I. V. Volovich, \textit{Quantum Theory
and Its Stochastic Limit\/} (Oxford University Press, London, in
press).
\bibitem{ref:vanHove}
L. van Hove, Physica \textbf{21}, 517 (1955).
\bibitem{ref:Davies}
E. B. Davies, Commun.\ Math.\ Phys.\ \textbf{39}, 91 (1974).
\bibitem{ref:TFD}
H. Umezawa, \textit{Advanced Field Theory: Micro, Macro, and
Thermal Physics\/} (American Institute of Physics, New York,
1993).
\end{references}
\end{document}